\documentclass[fleqn,usenatbib]{mnras}

\usepackage{newtxtext,newtxmath}

\usepackage[T1]{fontenc}

\DeclareRobustCommand{\VAN}[3]{#2}
\let\VANthebibliography\thebibliography
\def\thebibliography{\DeclareRobustCommand{\VAN}[3]{##3}\VANthebibliography}

\newcommand\gaia{\textit{Gaia }}

\newcommand\Pcut{$P_{\rm cut}$}

\newcommand\Teff{$T_{\rm eff}$}
\usepackage{graphicx}	% Including figure files
\usepackage{amsmath}	% Advanced maths commands
\usepackage{threeparttable} % to use table notes

\title[SB1 --- tidal circularization]
{Features of Gaia DR3 Spectroscopic Binaries \\ I. Tidal circularization of Main-Sequence Stars }

\author[D. Bashi et al.]{
Dolev Bashi,\thanks{E-mail: dolevbashi@gmail.com}
Tsevi Mazeh
and
Simchon Faigler\\
School of Physics and Astronomy, Tel Aviv University, Tel Aviv, 6997801, Israel\\
}

\date{Accepted XXX. Received YYY; in original form ZZZ}

\pubyear{2015}

\begin{document}
\label{firstpage}
\pagerange{\pageref{firstpage}--\pageref{lastpage}}
\maketitle

\begin{abstract}

Previous studies pointed out that many observed samples of short-period binaries display a cutoff period, \Pcut, such that almost all binaries with periods shorter than \Pcut\ have circular orbits.
This feature is probably due to long-term circularization processes induced by tidal interaction between the two stars of each binary. %\Pcut\ for different samples  was found at the range of $\sim3$--$15$ days, 
It seemed as if coeval main-sequence (MS) samples of open clusters display \Pcut\ that depends on the sample age.
Using  the unprecedentedly large sample of MS spectroscopic orbits recently released by \textit{Gaia} 
we have found that the \Pcut\ does not depend on the stellar age but, instead, varies with stellar temperature, decreasing {\it linearly} from $6.5$ day at $T_{\rm eff}\sim 5700$ K to $\sim 2.5$ day at $6800$ K.
%To further study the binary eccentricity 
\Pcut\ was derived by a new algorithm that relied on clear upper envelopes displayed in the period-eccentricity diagrams. 
Our \Pcut\ determines both the border between the circular and eccentric binaries and the location of the upper envelope. 
The results are inconsistent with the theory which assumes circularization occurs during the stellar MS phase, a theory that was adopted by many studies. 
The circularization has probably taken place at the pre-main-sequence phase,
as suggested 
already in 1989
by Zahn and Bouchet,
and later by Khaluillin and Khaluillina in 2011. % 
Our results suggest that the weak dependence of \Pcut\ on the cluster age is not significant, and/or might be due to the different temperatures of the samples.
%++++++++++++++++++++++++++++++++++
If indeed true, this has far-reaching implications for the theory of binary and exoplanet circularization, synchronization, and alignment.

\end{abstract}


%%%%%%%%%%%%%%%%% BODY OF PAPER %%%%%%%%%%%%%%%%%%

%========================
\section{Introduction}
\label{sec:intro}
%========================

The \gaia latest release of Non-Single Star catalogs 
\citep[][hereafter {\it NSS}]{NSS}
includes the orbits of $181\,327$ single-lined spectroscopic binaries (SB1), 
based on the radial velocities obtained by the space-mission RVS spectrograph 
\citep[][]{RVS_I_22,RVS_II_22,DR3_Katz}. 
The {\it NSS} SB1 catalog
is substantially larger than any previously-known catalog, and  
%For example, the SB9 catalog \citep[e.g.,][]{SB9} lists $\sim 4\,000$ orbits, while in a recent work based on the APOGEE project, \cite{Price-Whelan20} have identified $\sim 1\,000$ orbits.
%By increasing the known SB1s by two orders of magnitudes, 
therefore is a gold mine for investigating the statistical features of short-period binaries
\citep[e.g.,][]{DuquennoyMayor91,duchene13,TorresLathamQuinn21}, like 
%the period-eccentricity relation \citep[e.g.,][]{Mazeh2008,jorissen09}, 
the frequency of binaries as a function of the primary mass \citep[][]{raghavan10,troup16, moe17} and mass-ratio distributions \citep[e.g.,][]{mg92,boffin12, boffin15,shahaf17}.

%The \gaia catalog includes the Keplerian elements of the orbits, but not the \gaia RVs and their epochs themselves. Naturally, some erroneous orbital solutions are probably hidden in the catalog. used two external sources to validate the orbit and devised a clean but still large Gaia SB1 sample of $91\,740$  orbits.  The sample differs from the parent sample, for example, by the absence of --- physically unlikely and hence presumably spurious --- short-period binaries with high eccentricity. 

%These features might have profound implications on our understanding of binary formation and evolution \citep[e.g.,][]{verbunt95,bate97,bate02,harada21}, and therefore were intensively discussed in the past \citep[e.g.,][]{duchene13,shahaf19}, based on the relatively small samples available then. The field is now open for new detailed studies based on the much-larger \gaia-SB1 catalog. 

This work utilizes the {\it NSS} catalog 
to follow the tidal circularization of short-period binaries. 
%study the circularized orbits of the short-period \gaia binaries.
%We identify periods that separate circular from eccentric binaries, and follow the dependence of these periods on stellar temperature. 
Based on theoretical work, we expect short-period binaries ($\lesssim 3$ -- $10$ day) to be circularized \citep[e.g.,][]{kopal56,MayorMermilliod84,giuricin84, zahn08} by the tidal interaction between the two components, whose strength is a strong function of the binary separation and primary radius \citep[e.g.,][]{zahn75, zahn77,zahn89}. This expectation caused \cite{NSS} to question the validity of the very short-period binaries, with periods shorter than one day and large eccentricities. We, therefore, rely on the work of \cite{bashi22}, who
confronted the {\it NSS} SB1 orbits with the LAMOST and GALAH RV databases,
and constructed a clean sample of $91\, 740$ \gaia SB1 orbits. This sample is the basis of our period-eccentricity relation.  
%th dependence of the interaction strength on the stellar parameters,  
%an argument that was one of the reasons to construct the  clean sample.  

Observational evidence for tidal circularization, based on small samples of spectroscopic binaries,  was already pointed out by \cite{MayorMermilliod84} and
 \cite{MathieuMazeh88}. They considered samples of coeval spectroscopic binaries and showed that each sample displays a circularization "cutoff period", \Pcut,
%$P_{\rm cutoff}$ 
out to which most binaries have circular orbits, while most binaries of longer periods have eccentric orbits \citep[see also][]{mathieuMeibomDolan04}. 
The idea behind the cutoff period is the strong dependence of the tidal interaction on the binary separation and therefore on the binary period. Binaries with periods longer than \Pcut\ were not circularized,  keeping their primordial eccentricity. In this model, coeval binary samples should display a discontinuity jump of the eccentricity at \Pcut, which should depend on the sample age --- the older the sample the longer \Pcut\ 
\citep[e.g.,][]{MathieuMazeh88,mathieuMeibomDolan04,geller08, geller10,geller12,nine20,geller21,zanazzi22}.

Another avenue to observationally study the tidal circularization was taken by \cite{NorthZahb03} and \cite{MazehTamuzNorth06},  and recently by the seminal works of \cite{EylenWinnAlbrecht16} and \cite{JustesenAlbrecht21}; see also \cite{zanazzi22}. They considered samples of eclipsing binaries (EB) --- those of OGLE LMC \citep{LMC03}, {\it Kepler} \citep{slawson11} and {\it TESS} \citep{TESS15}, obtaining the projected eccentricity of those binaries from the timing of the secondary eclipse
\citep{sterne40}.
%and derived \Pcut, or the cutoff 
%
These studies used the fact that a light curve of an EB allows for deriving the stellar radii relative to the binary separation, and therefore the 
ratio between the sum of radii of the two components and the binary separation, a ratio that directly determines the tidal circularization effectiveness \citep{NorthZahb03}. Thus, the   
analysis of EB samples provides a cutoff ratio that divides circular from eccentric orbits. \cite{JustesenAlbrecht21} derived the cutoff ratio for binaries with different temperatures. Such a ratio is not available for spectroscopic binaries, and therefore one can study only the cutoff period.

When studying spectroscopic samples, 
 \cite{meibom05}
%considering the SB ? of the open cluster M35, 
suggested a more composite approach for determining the cutoff period of a binary sample. They considered the eccentricity as a piece-wise function of the period, $P$, with zero eccentricity for
$P<P_{\rm cut}$ and a smooth rising asymptotic function that goes up to eccentricity close to unity  for 
$P>P_{\rm cut}$.
The idea behind the \cite{meibom05} approach is that even binaries with $P>P_{\rm cut}$ are subject to the circularization processes that decreased the primordial eccentricity, but were not strong enough to make the orbit circular. 
The reduced eccentricities of those binaries depend on the primordial orbital period and eccentricity and the strength of the tidal interaction.
The asymptotic function reflects the present characteristics of the whole binary sample and not only the very short-period circularized binaries (see this approach applied, for example, by \cite{nine20} and with a different version by \cite{zanazzi22}).

Following \cite{Mazeh2008} \citep[see also, for example,][]{SB9}
we consider here a piece-wise asymptotic function that 
presents an upper envelope for the eccentricity as a function of the period for a given sample.  
The function marks
the edge of the populated part of the period-eccentricity plane. 
For a given period, there are almost no binaries with eccentricity larger than the asymptotic-function value at this period, while there is a high probability of finding binaries with eccentricities smaller than that value. In short, we view the period-eccentricity diagram as divided into two regimes, separated by an upper-envelope asymptotic function. 
We use  \cite{Mazeh16} algorithm to find the best upper envelope by a maximum-likelihood approach, given a sample of spectroscopic binaries. Similar to \cite{meibom05}, we define the cutoff period, \Pcut, as the period at which the upper envelope cuts the eccentricity axis, i.e., $e=0$. 
In this approach, \Pcut\ denotes both  the border between the circular and eccentric binaries and the location of the upper-envelope base. 

The large cleaned \gaia sample of \cite{bashi22} allows us to apply our technique to a few sub-samples of the \gaia SB1s. We concentrate on the binaries with main-sequence (MS) A-, F- and G-primaries, identified by their position on the \gaia Colour-Magnitude 
Diagram (CMD). Similar to the \cite{JustesenAlbrecht21} analysis, we divided the sample into temperature bins, and derived their corresponding \Pcut\ independently, finding a clear dependence on the stellar temperature of each bin.  

Section~\ref{sec:sample} details the \gaia sample used, including removing binaries with unconstrained  
eccentricities. Section~\ref{sec:envelope} presents our algorithm to obtain the best parameters of the upper envelope, given a binary sample, and Section~\ref{sec:Simulations} tests our algorithm for six simulated binary samples.
Section~\ref{sec:p0_temp} presents our main result --- the dependence of \Pcut\ on the primary temperature of each sub-sample. 
Section~\ref{sec:tidal-theory} shows that
the results are inconsistent with the theory which assumes circularization occurs during the stellar MS phase.
Section~\ref{sec:discussion} summarizes our results and discusses their possible implications.

%========================
\section{The sample}
\label{sec:sample}
%=======================

%--------------------------------------------------------------
 \begin{figure}
	\includegraphics[width=9cm]{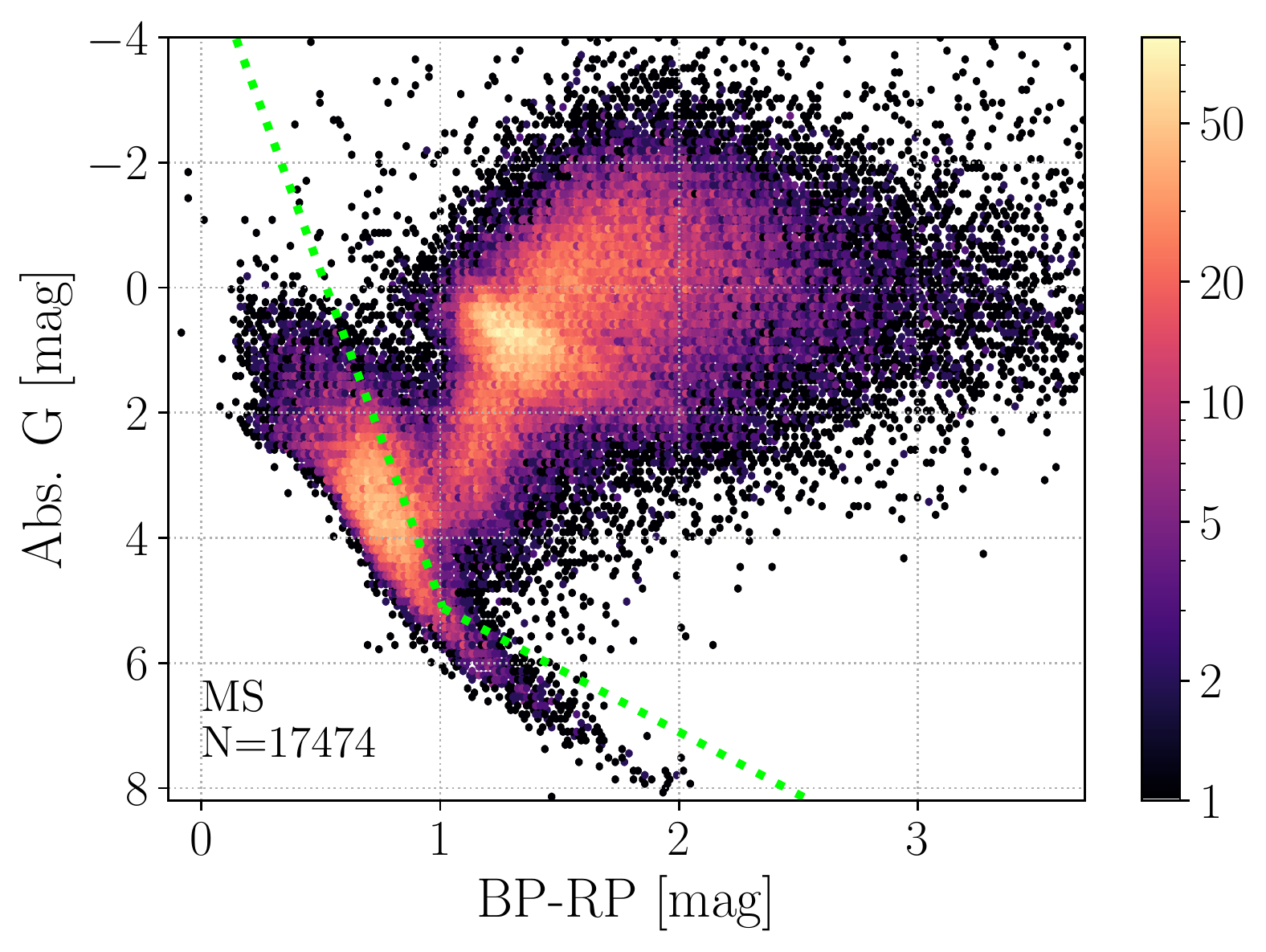}
    \caption{Colour-Magnitude Diagram (CMD) of the clean \gaia sample of \protect\cite{bashi22} binaries. The colour scale represents the number of binaries falling into each  hexagonal bin. The dashed line (see text) delineates our border of the MS binaries.}
    \label{fig:CMD}
\end{figure}
%------------------------------------------------------------

We wish to study the eccentricity distribution of the \gaia binaries, as a function of the stellar temperature in particular. We, therefore, consider a uniform sample of binaries with unevolved primaries, of derived effective temperature, and orbits with well-constrained eccentricity. Such a reduced sample is composed in this section. 

\subsection{The main-sequence binaries}
%---------------------------------------

We first cross-matched the clean sample of \cite{bashi22} with the TESS Input Catalog\footnote{https://tess.mit.edu/science/tess-input-catalogue/} \citep[TIC; ][]{TIC8, TIC8_2}, yielding  $89\, 742$ SB1s. 
Fig.~\ref{fig:CMD} shows these binaries on the \gaia Color-Magnitude Diagram (CMD),
%the position of the of \cite{bashi22} that are listed in the On the CMD, 
clearly displaying the difference between the MS and the evolved stars. To identify the MS binaries we used the limits
%---------------------------------------------------------------
\begin{equation}
  \mathrm{MS}:\begin{cases}
    -5.5+10.5(G_{\mathrm{BP}}- G_{\mathrm{RP}}) < G, & \mathrm{if~} G_{\mathrm{BP}}- G_{\mathrm{RP}} \leq 1 \ ,\\
    3.1+2(G_{\mathrm{BP}}- G_{\mathrm{RP}}) < G, & \mathrm{if~} G_{\mathrm{BP}}- G_{\mathrm{RP}} > 1 \ ,
  \end{cases}
\end{equation}
%---------------------------------------------------------------
which left us with $17\, 474$ binaries.

%----------------------------------
\subsection {Removing orbits with unconstrained  eccentricity}
\label{sec:e-p MS sample}
%----------------------------------

%---------------------------------------------------
 \begin{figure}
	\includegraphics[width=8.5cm]{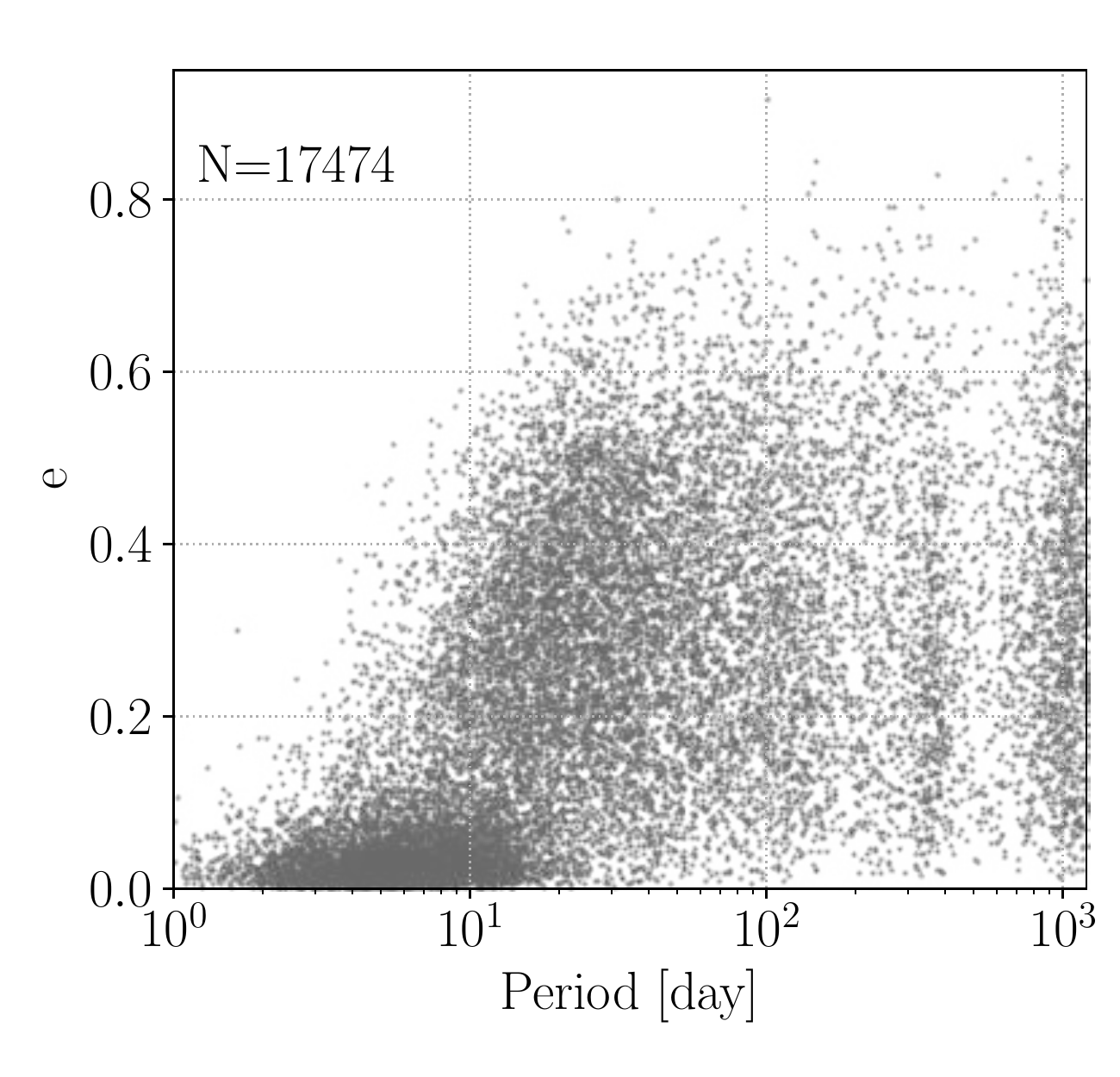}
    \caption{Period-eccentricity diagram for the cleaned MS SB1 sample.}
    \label{fig:per_ecc_all_MS}
\end{figure}
%----------------------------------------------------

Fig.~\ref{fig:per_ecc_all_MS} shows the period-eccentricity ($P$-$e$) diagram for the MS-reduced sample. Two features are clearly seen:

%---------------------------------------------------------
\begin{itemize}
\item[--] An upper envelope that rises from $e\sim 0.1$ at period of $P\sim 3 $ day towards $e\sim 0.8$ at period of $P\sim 100 $ day.
\item[--] A high concentration of binaries centered at $P\sim 5$ day and low eccentricity of $\sim 0.03$. 
\end{itemize}
 %--------------------------------------------------------------

To check which of these low eccentricities are real \citep[see also a discussion by][]{bashi22}, we used here the \cite{LucySweeney71} approach \citep[see also][in the context of the eccentricity of planetary orbits]{hara19}, who suggested that the actual eccentricity uncertainty is larger than the one derived by the regular analysis.  Thus, \cite{LucySweeney71} suggested that many derived eccentricities of their time might be spurious.

%---------------------------------------------------
 \begin{figure}
	\includegraphics[width=8.6cm]{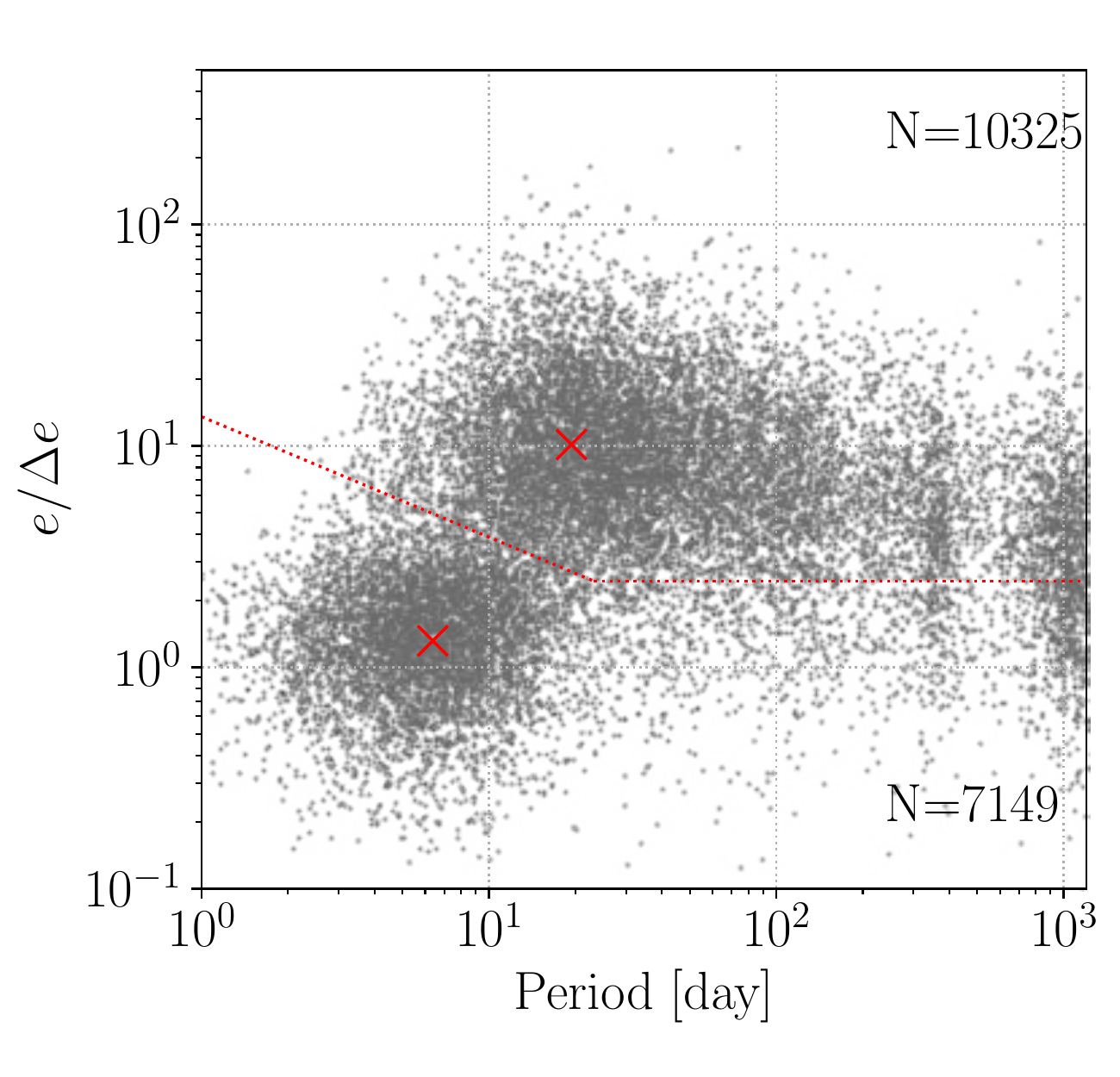}
    \caption{Ratio between the derived eccentricities and their corresponding uncertainties as a function of the orbital period. Red dashed horizontal line marks the $e/\Delta e = 2.45$ threshold of \protect\cite{LucySweeney71}.}
    \label{fig:ecc_dee_MS}
\end{figure}
%----------------------------------------------------

We, therefore, plot in Fig.~\ref{fig:ecc_dee_MS} the ratio between the derived \gaia eccentricities and their corresponding uncertainties as a function of the orbital periods. Two emerging clusters can be seen for binaries with orbital periods shorter than $100$ days. Using K-means clustering \citep{sklearn}\footnote{https://scikit-learn.org/stable/modules/generated/sklearn.cluster.KMeans.html}, we found the center of the bottom cluster at ($P,e/\Delta e$) = ($6.37$, $1.32$), while the upper-cluster center is at ($19.38$, $10.15$). Using the \cite{LucySweeney71} $5\%$ level of significance for testing the hypothesis that $e=0$, we found most binaries in the lower cluster to be under the threshold of $e/\Delta e = 2.45$.
To define a sample of binaries with well-constrained eccentricities, we drew a line perpendicular to the line connecting the two cluster centres through the middle of the gap between the two clusters.
 %We ignored binaries under this line or under the $e/\Delta e = 2.45$ limit. 
This has left us with a sample of $10\,325$ binaries with at least $95\%$ confidence for non-circular eccentricity, and $7\,149$ binaries with unconstrained eccentricities.

\begin{figure*}
	\includegraphics[width=16cm]{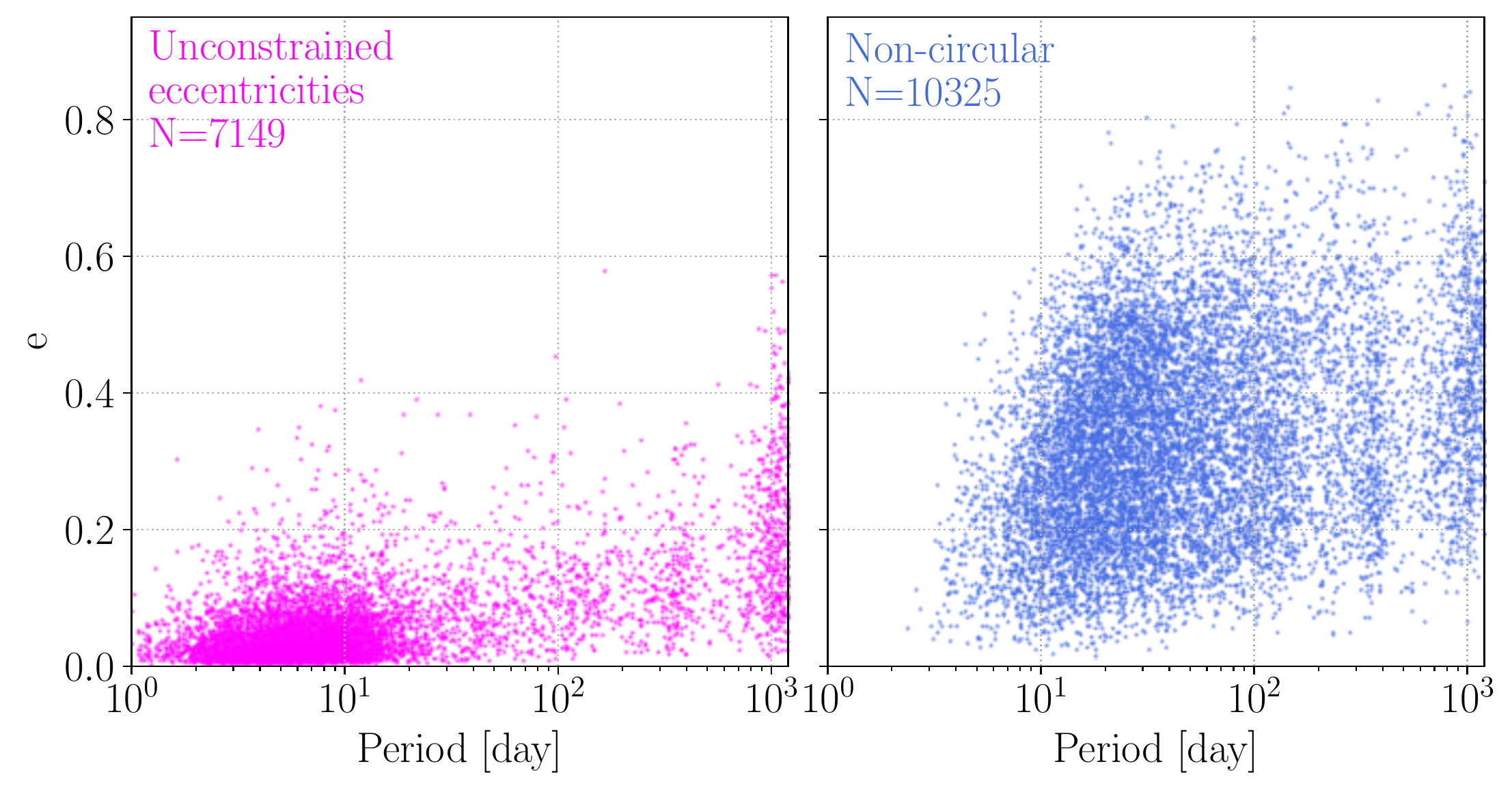}
    \caption{Period-eccentricity diagram for the cleaned SB1 sample of MS unconstrained eccentricities (left) and non-circular (right) binaries.}
    \label{fig:per_ecc_MS}
\end{figure*}

Fig.~\ref{fig:per_ecc_MS} presents the two sub-samples.  
As can be clearly seen, the dense area of near-circular binaries is populated by the unconstrained eccentricity sources only. 

%---------------------------------------------------
%----------------------------------------------------

%-------------------------------------
\begin{figure*}
	\includegraphics[width=16cm]{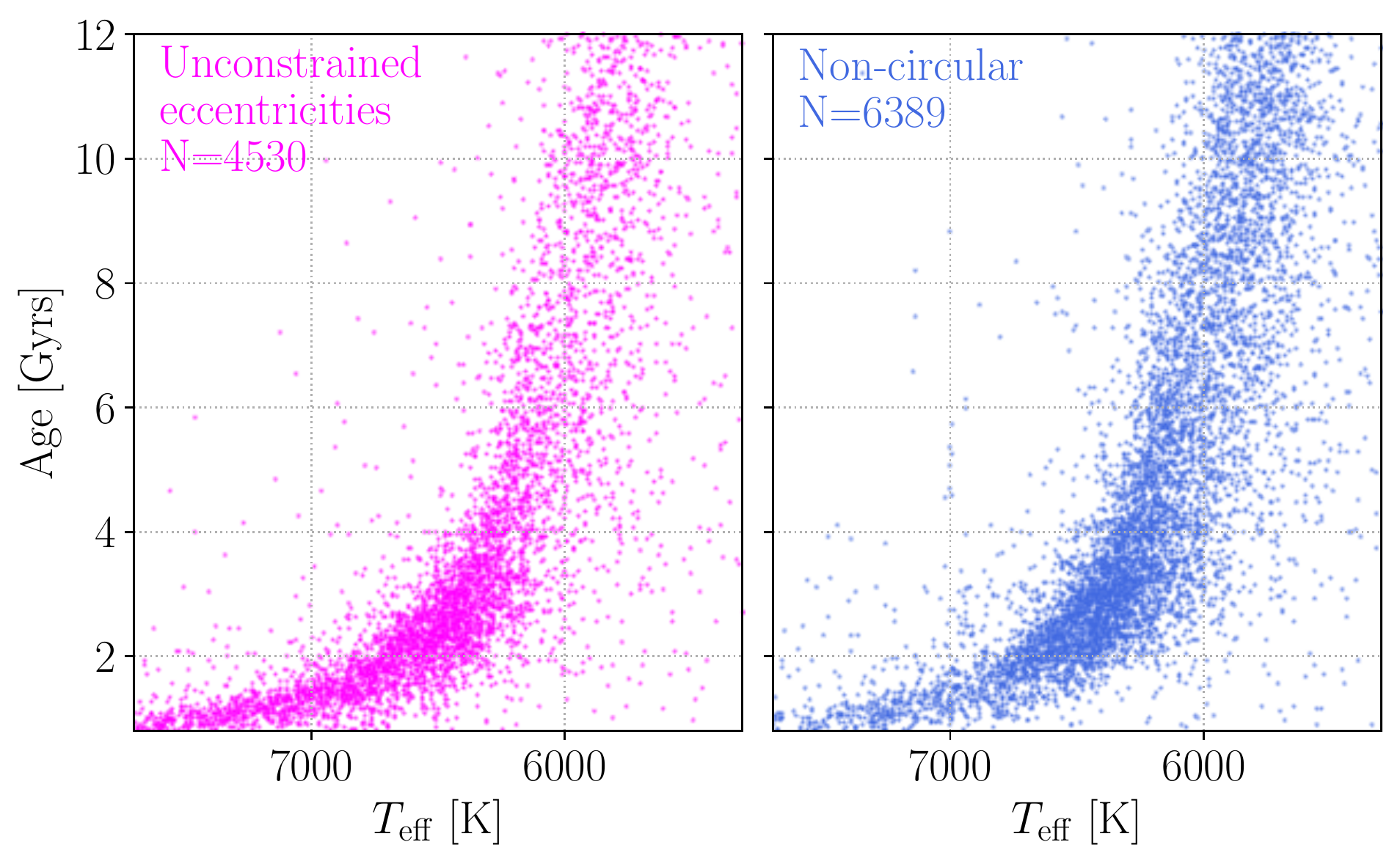}
    \caption{Stellar Age-effective temperature scatter plot of the unconstrained eccentricities sample (left) and non-circular (right) samples. A similar trend between the two parameters is evident in both panels.}
    \label{fig:AgeTeff}
\end{figure*}
%--------------------------------------

Next, we wish to test whether the clear distinction between the two sub-samples of Fig.~\ref{fig:per_ecc_MS} is a result of an 
age difference, as one might argue that the nearly circular binaries are old MS binaries that had already completed their tidal circularization. We, therefore, plot in Fig.~\ref{fig:AgeTeff} the age-effective temperature dependence of each of the two sub-sample. 
The left panel includes a sample of $4\,530$ binaries with unconstrained eccentricities for which  \gaia FLAME stellar effective temperature and age 
(\texttt{age\_flame}) are available. Similarly, the right panel includes a sample of $6\,389$ sources with non-circular solutions. Overall, a strong correlation between effective temperature and age is evident in both panels, as expected for MS stars, with no apparent age difference. 

In what follows, we concentrate on analysing the sub-sample of $10\,325$ binaries with well-constrained eccentricities, deriving the dependence of \Pcut\ on the stellar temperature. We will also show that \Pcut\ does not display a dependence on stellar age.

%\textcolor{red}{I believe this line should be revised:} In deriving the upper envelope we ignored the almost circular orbits, with $e<0.1$, and the long-period binaries, with $P>30$, as delineated in the figure. Thus, we were left with only $2\, 667$ binaries. 

%============================
\section{Fitting an Upper Envelope to the period-eccentricity diagram}
\label{sec:envelope}
%============================

The right panel of Fig.~\ref{fig:per_ecc_MS} clearly shows that 
the binary eccentricities do not present a clear cutoff period that separates the circular and the eccentric binaries, but, instead, show a gradual rise toward high values.
Furthermore, 
the eccentricities are not concentrated around an upper envelope 
but are spread out between almost zero eccentricity and the upper bound. 
 The figure suggests that the base of the envelope can serve as a better definition for the cutoff period. 
 Any circularization analysis should try to account for this feature.  

To better characterize  the $P$-$e$ diagram we define an upper envelope as
%
%---------------------------------------------------
\begin{equation}
%f = 1 - \mathrm{exp}(-(\mathcal{P} - \mathcal{P}_0)/\tau)\ ,
 f(P) =\begin{cases}
    1-\bigg(\frac{P_{\mathrm{cut}}}{P}\bigg)^{1/\tau} & \quad  \quad \quad P_{\mathrm{cut}} < P  \ ,\\
    0 & \quad  \quad \quad
 P_{\mathrm{cut}} \geq P \ ,
  \end{cases}
  \label{eq:e_p_env}
\end{equation}
%---------------------------------------------------
%
where \Pcut\ is the cutoff period 
\citep[see also][]{zanazzi22} and $\tau$ is a dimensionless parameter that determines the slope of the envelope. The function gets a value of zero at \Pcut\  for any value of $\tau$, and asymptotically rises to unity as $P$ gets substantially larger than \Pcut. 
%For a certain value of eccentricity and \Pcut, $\tau$ is also a measure of what should be the expected period that meets the envelope curve. 
%
Note that the function is simple with only two free parameters --- \Pcut\, and $\tau$.
\Pcut\ defines both the separation between the circular and eccentric orbits and the location of the upper envelope.

\begin{figure}
        \includegraphics[width=8.5cm]{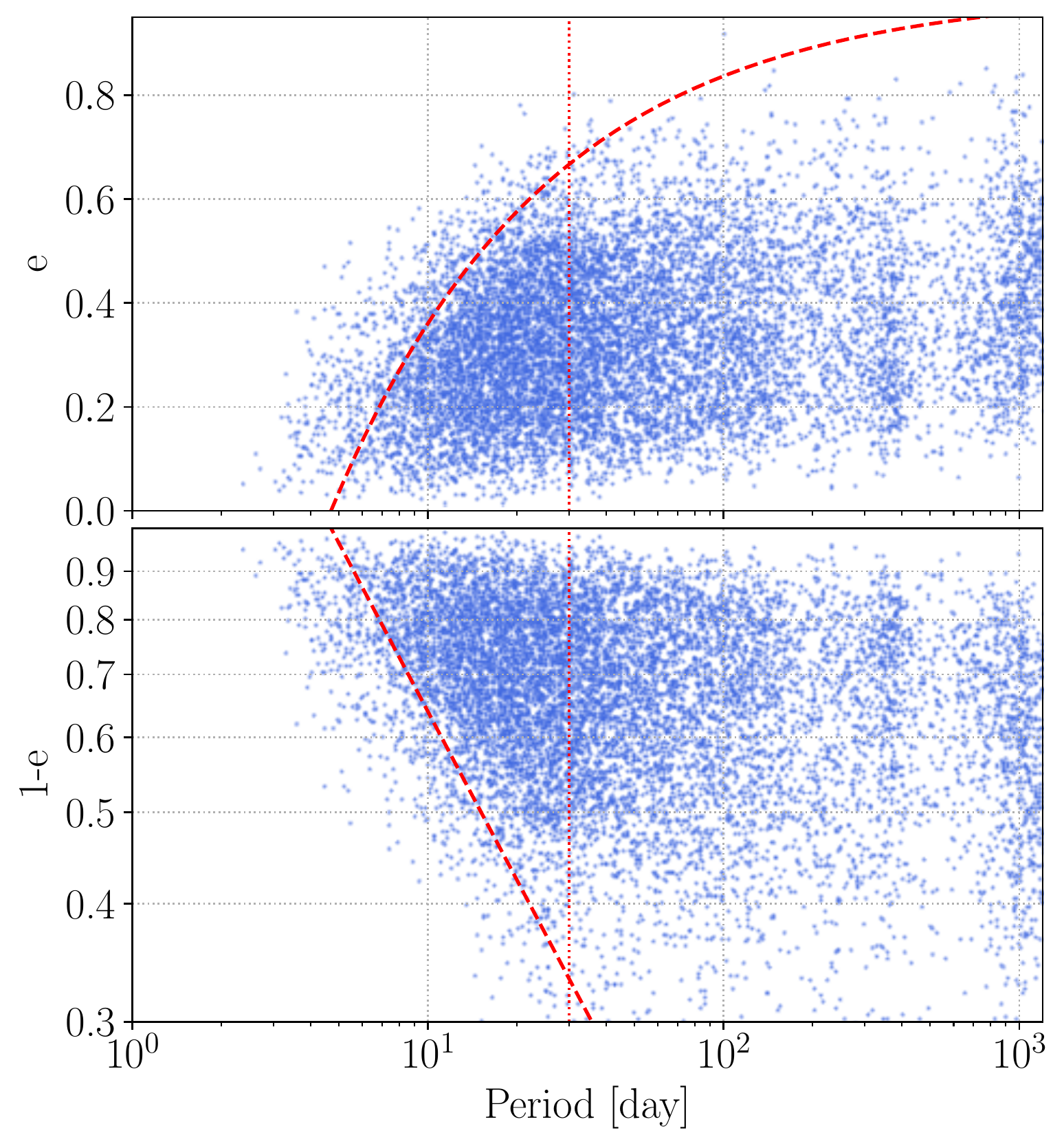}
    \caption{Top: Log period -- eccentricity 
      diagram for the cleaned SB1 sample of MS binaries, after removing orbits with unconstrained eccentricities (see text).
    %based on the \protect\cite{LucySweeney71} limit.
    The red dashed line marks our upper envelope best-fit model, based on binaries with an orbital period shorter than $30$ days, marked by a vertical dotted line (see Section \ref{sec:envelope} for more details).
     Bottom: The same sample plotted on the ($\log P$, $\log (1-e)$) plane. The same upper envelope is now a straight-line lower envelope, with a slope of $\tau$.
    }
    \label{fig:per_ecc_MS_env}
\end{figure}

To get a sense of the role of $\tau$, one can  plot
(see the bottom panel of Fig.~\ref{fig:per_ecc_MS_env})
the analysed sample in the ($\log P$, $\log (1-e)$) plane. In this plane, the upper envelope turns into  a straight-line lower envelope with a slope of $-1/\tau$. 
%
%The bottom plot demonstrates the role of $\tau$ --- controlling the slope of the upper (lower) envelope. 
%
Another way to follow the meaning of $\tau$ is 
to consider, for example, 
$P_{0.8}(P_{\rm cut}, \tau)$ --- the period for which the upper envelope
gets the value of $e=0.8$. It is easy to show that 
$
    P_{0.8}=5^{\tau}P_{\rm cut}.
$
In other words, $\tau$ measures by how much we have to increase $P$, relative to $P_{\rm cut}$, for the upper envelope to get a value of $e=0.8$.

We then follow \cite{Mazeh16} and use
%a maximum likelihood with 
a modified Fermi function to describe an assumed probability density function (PDF) of binaries above and below the upper envelope in the ($\log P$, $e$) plane.
The PDF converges to zero above the envelope, and to a positive constant below it.  The transition region between the two parts is along the envelope, with a width characterized by a parameter $\delta$.
%as defined in equation~\ref{eq:e_p_env}. 
%
The probability density at a point $(\log P,e)$ is
a function of its distance $d(\log P,e;\, P_{\mathrm{cut}},\tau)$ to the envelope curve along the $\log P$ axis and can be expressed as:
\begin{equation}
d = \log P - \log f^{-1}(e) = \log \Bigg[\frac{P}{P_{\mathrm{cut}}} (1 + e)^{\tau}\Bigg]\ ,
\label{eq:dist}
\end{equation}
where $f^{-1}(e)$ is the inverse function of $f(P)$ defined in equation~\ref{eq:e_p_env}. The distance $d$ is positive for points to the right and below the envelope and negative on the other side.
%and $\delta$ is the transition width. 
The PDF is 
%---------------------------------------------------------------
\begin{equation}
\mathcal{F}_{\mathrm{PDF}}(P,e;\,{P_{\mathrm{cut}}},\tau,\delta) =  \mathcal{A}\frac{1}{1+ \mathrm{exp}(-d/\delta)}\ ,
\label{eq:PDF}
\end{equation}
%---------------------------------------------------------------
where the constant $\mathcal{A}$, in units of
$1/[\log(P/1\, {\rm day)}\, e]$, is defined such that the 2D integral
over the ($\log P$, $e$) plane, using the three parameters --- ${P}_{\mathrm{cut}},\tau,\delta$, equals unity. To get a sense of this function, one can see that when $d=0$ the PDF value is simply 
$\mathcal{A}/2$, while at $d=\pm \delta$ it changes between 
$0.27\mathcal{A}$ and $0.73\mathcal{A}$. 

Obviously, the probability density function is of a statistical nature. One can find binary systems above the upper envelope, either due to erroneous measurements or invalid orbits or because of some special astrophysical circumstances, like a high primordial eccentricity or a distant faint companion that pumped eccentricity into the spectroscopic orbit \citep[e.g.,][]{MazehShaham79,mazeh90,jha00}. The advantage of our approach is that 
for large enough samples the derivation of the envelope is only slightly sensitive to these 'outlier' binaries, as the fit is to the population as a whole. On the other hand, the results are not biased by the number of binaries in the sample.

For each set of parameters of the PDF --- 
${P}_{\rm cut}$, $\tau$ and  $\delta$,
we derive the likelihood of a given sample of $N$ binaries, 
as
%---------------------------------------------------------------
\begin{equation}
\mathcal{L} = \prod_{i=1}^{N} \mathcal{F}_{\mathrm{PDF}}({P}_i, e_i; {P}_{\rm cut},\tau,\delta) \ .
\label{eq:likelihood}
\end{equation}
%---------------------------------------------------------------
%
where $e_i$ and $P_i$ are the eccentricity and period of the $i-$th binary.

%We used the following uninformative priors on the free parameters: $\log P_{\mathrm{cut}} \sim \mathcal{U}(-1,1)$; $\tau \sim \mathcal{U}(0,4)$ and $\delta\sim \mathcal{U}(0.01,0.3)$ for the MS sample,
%and $\log P_{\mathrm{cut}} \sim \mathcal{U}(1,2)$; $\tau \sim \mathcal{U}(0,2)$; $\delta\sim \mathcal{U}(0.01,0.3)$ for the Clump sample.

We applied our algorithm to the sample of the \gaia well-constrained eccentricities discussed above, using only a reduced sub-sample of 
%add to the figure a fitted upper envelope using the sub-sample 
$4\,430$ binaries with orbital periods shorter than 30 days.
This was done because the density of the binaries with periods larger than 30 days is clearly not constant, contrary to the assumption of our algorithm. 
%(see next section for details). }
To find the upper envelope of the reduced MS sample of Fig.~\ref{fig:per_ecc_MS} we ran an MCMC routine with $50$ walkers and $10^4$ steps, with uninformative priors on the free parameters: $\log P_{\mathrm{cut}} \sim \mathcal{U}(-1,1)$; $\tau \sim \mathcal{U}(0,4)$ and $\delta\sim \mathcal{U}(0.01,0.3)$.
We used the Python \texttt{emcee} package \citep{foreman-Mackey2013} to find the parameter values (and their uncertainties) that maximize the sample likelihood. 

Fig.~\ref{fig:per_ecc_MS_env} shows our best-fit upper envelope
%derived for the reduced binary sample \textcolor{red}{of $4\,430$ sources with an orbital period shorter than 30 days} 
with $P_{\mathrm{cut}} = 4.67~\mathrm{day}$, $\tau = 1.70$ and $\delta=0.09$.
The bottom panel of the figure displays 
the analyzed sample in the ($\log P$, $\log (1-e)$) plane. In this plane, the upper envelope turns into  a straight-line lower envelope with a slope of $-1/\tau$,  showing how $\tau$ controls the slope of the upper (lower) envelope.

%--------------------------------------------------------------
%--------------------------------------------------------------

%===================================
\section{Testing our algorithm on simulated binary samples }
\label{sec:Simulations}
%===============================
To test our upper-envelope algorithm, we performed a simplified simulation of six binary samples with different 
%effectiveness of 
circularization efficacy coefficients, all starting with the same flat distribution of eccentricity and log-period. 
%The results of the simulation are presented in 

The binaries were numerically evolved with \cite{hut81} \cite[see also][]{TerquemMartin21} equations \citep[][]{zahn75,zahn77,zahnBouchet89,zahn89} for stars with convective envelopes:
%----------------------------------------------------
\begin{equation}
\begin{aligned}
%& \frac{de}{dt}= -e\,B{P^{-16/3}},\ \ {\rm and} \\
%&\frac{dP}{dt}=-3e^2\,{B}{P^{-13/3}}\ .
& \frac{de}{dt}= -\frac{18}{7} B \Big(\frac{P}{1~ \rm{day}}  \Big)^{-16/3} \frac{e}{(1-e^2)^{13/2}}\Bigg[1+ \frac{15}{4}e^2 + \frac{15}{8}e^4 + \frac{5}{64}e^6 \\ 
& \ \ \ \ \ \ \ \ \  - \frac{11}{18}(1-e^2)^{3/2} \Big( 1 + \frac{3}{2}e^2 + \frac{1}{8}e^4 \Big) \Bigg],\ \ {\rm and} \\
&\frac{d}{dt} \Big(\frac{P}{1~ \rm{day}}  \Big)=-\frac{6}{7} B \Big(\frac{P}{1~ \rm{day}}  \Big)^{-13/3}\frac{1}{(1-e^2)^{15/2}} \Bigg[1 
 + \frac{31}{2}e^2 + \frac{255}{8}e^4 \\ 
& \ \ \ \ \ \ \ \ \ \ \ \ \ \ \ \ \ \ \ \ \   + \frac{185}{16}e^6 + \frac{25}{64}e^8  -(1-e^2)^{3/2} \Big( 1 + \frac{15}{2}e^2 + \frac{45}{8}e^4 + \frac{5}{16}e^6\Big) \Bigg]\ .
\end{aligned}
\label{eq:tidal_sim}
\end{equation}
%----------------------------------------------------
%
We chose the constant $B$ to control the circularization efficacy over the binary lifetime, which we define as going from  $t=0$ to $t=1$, 
where $t$ is unitless. 
%So is $P$, defined as the orbital period in units of one day.

%--------------------------------------------------------------
\begin{figure*}
	\includegraphics[width=18cm]{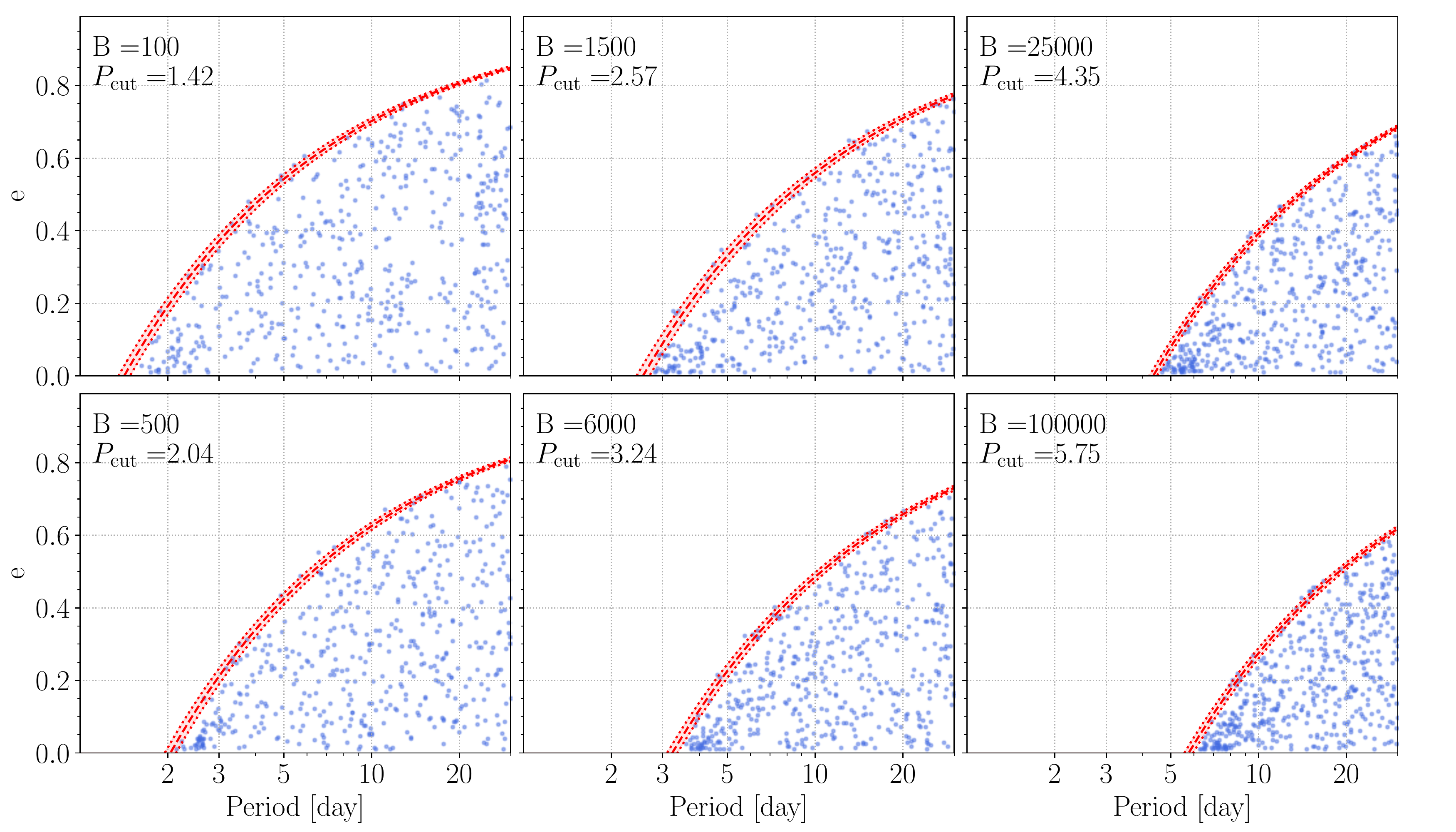}
    \caption{Simulated binary samples of $500$ binaries each that were integrated through the circularization process following equations~\ref{eq:e-p_int}. The red lines show the upper envelopes, while the red area marks the transition region of $\pm \delta$ width along the envelope. 
    }
    \label{fig:per_ecc_sim}
\end{figure*}
%--------------------------------------------------------------

Note that we ignore the changes in the stellar parameters, like radius and structure, and the stellar rotation, which must account for the varying angular momentum of the orbital motion. Furthermore, 
equations~\ref{eq:tidal_sim} are a simplified version of the actual tidal interaction, neglecting the dependence on the binary mass ratio, for example. Nevertheless, they are good enough for our purpose here \citep[see, for example, the discussion of][]
%[who argue that the  terms of higher of ]
{Terqem21}.

We note that if we keep only the leading order of the eccentricity in equations~\ref{eq:tidal_sim} (i.e., $e<<1$), we get: 
%----------------------------------------------------
\begin{equation}
\begin{aligned}
& \frac{de}{dt}= -eB \Big(\frac{P}{1~ \rm{day}}  \Big)^{-16/3}, \ \  {\rm and} \\ 
&\frac{d}{dt} \Big(\frac{P}{1~ \rm{day}}  \Big)=-\frac{57}{7}e^2 B \Big(\frac{P}{1~ \rm{day}}\Big)^{-13/3}.
\end{aligned}
\label{eq:tidal_approx}
\end{equation}
%----------------------------------------------------
%
The approximated equations allow one to derive the eccentricity as a function of the period for a given binary
%We write ------------------------------------------------------
%\begin{equation}
%\begin{aligned}
%&\frac{dP}{de}=3e\, P,\\ 
%&\frac{dP}{P}=3e\,de, 
%\end{aligned}
%\label{eq:e-p}
%\end{equation}
%---------------------------------------------------------------
%the solution of which is
%---------------------------------------------------------------
\begin{equation}
\begin{aligned}
%&ln P\big]^P_{P_0}=\frac{3}{2}e^2\big{]}^e_{e_0}\,  \\
&\ln(P/P_0)=\frac{57}{14}(e^2-e^2_0)\ ,
\end{aligned}
\label{eq:e-p_int}
\end{equation}
%--------------------------------------------------------
where $P_0$ and $e_0$ are the period and eccentricity at the starting point of the circularization track.
This equation does not predict the amount of change the eccentricity and period will go through in a given timespan but instead draws an evolutionary trajectory in the ($\log P$, $e$) plane.

%-----------------------------------------------------
\begin{table}
\caption{
Best-fitted values for \Pcut, $\tau$, $\delta$ of our simulated binary sample of $500$ binaries for six bins of circularization parameter $B$}
%{\color{red} maybe the range of temperature}
%-------------------------------------
\begin{center}
\begin{tabular}{r c c c }
\hline
B\ \ \ \ & \Pcut~[day] & $\tau$ & $\delta$ \\
\hline
$100$ & $1.42\pm 0.05$ & $1.61\pm 0.05$ & $0.020\pm 0.007$\\
$500$ & $2.04\pm 0.05$ & $1.62\pm 0.04$ & $0.020\pm 0.005$\\
$1500$ & $2.57\pm 0.07$ & $1.66\pm 0.04$ & $0.022\pm 0.005$\\
$6000$ & $3.24\pm 0.08$ & $1.70\pm 0.05$ & $0.018\pm 0.005$\\
$25000$ & $4.35\pm 0.08$ & $1.67\pm 0.04$ & $0.013\pm 0.003$\\
$100000$ & $5.75\pm 0.12$ & $1.72\pm 0.05$ & $0.015\pm 0.004$\\
\hline
\end{tabular}
\end{center}
\label{tab:P0-B}
\end{table}
%----------------------------------

Table \ref{tab:P0-B}, as well as the six panels of Fig.~\ref{fig:per_ecc_sim}, present the results of the integration through the "lifetime" of the samples, for six different values of the circularization parameter, or alternatively for different lifetimes; these values span a range of three orders of magnitudes. 
Each sample contains $500$ binaries of the same lifetime and circularization effectiveness $B$. Other simulations, not shown in the figure, showed that the best-fit obtained parameters do not depend on the number of binaries in the sample; only the uncertainties of the parameters do.

The panels clearly show that the separation between the circular and eccentric orbit is not a straight line but has a curved shape, supporting our approach that uses an upper envelope to characterize the period-eccentricity relation. Those envelopes are included in the figure, together with  their corresponding cutoff periods.  

\begin{figure}
	\includegraphics[width=8.5cm]{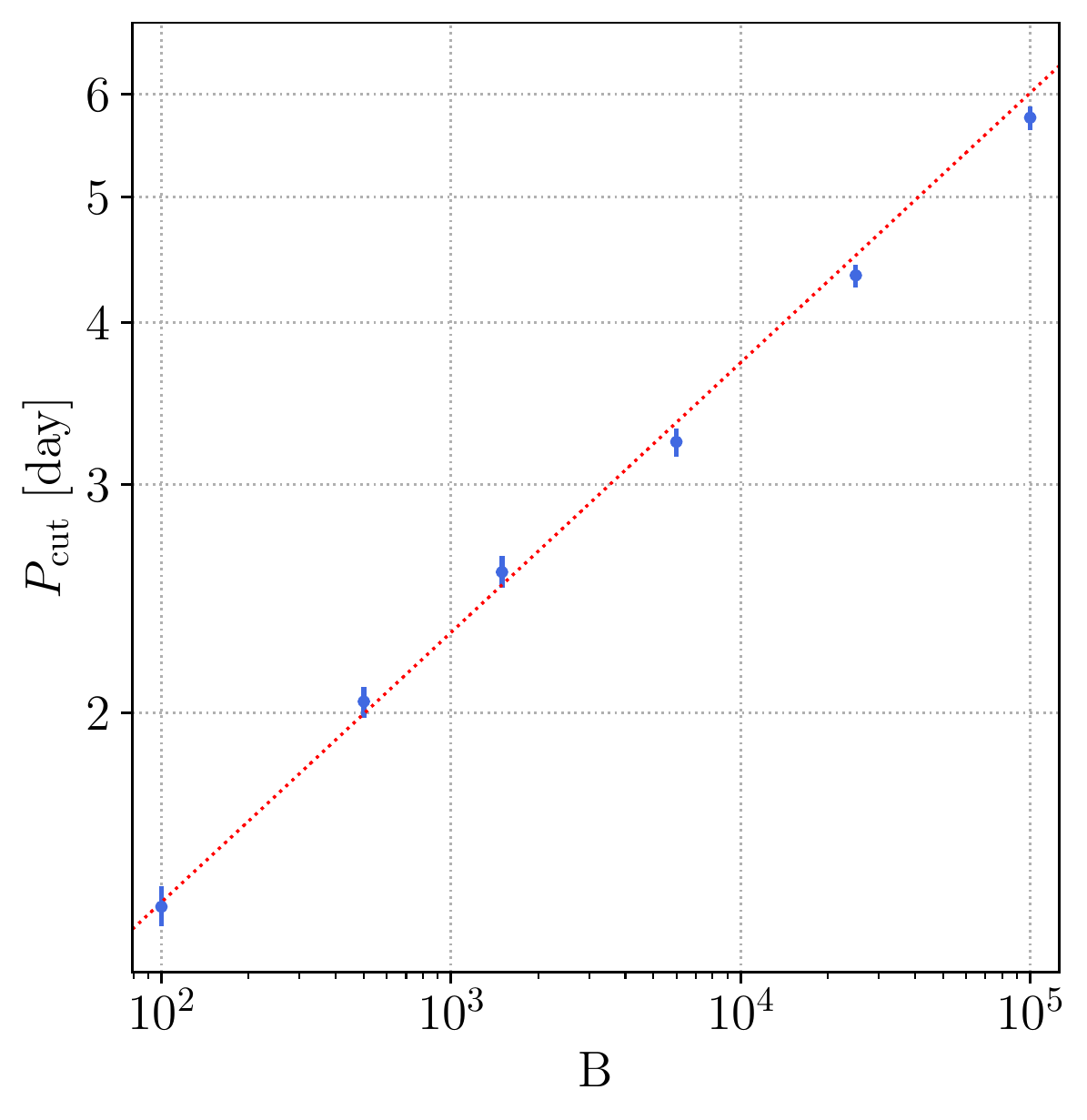}
    \caption{\Pcut\ as function tidal dissipation effectiveness $B$ for six simulated samples. Red-dotted line,
     $\log P_{\rm cut} = (0.198 \pm 0.011)\, \log (B) - (0.231 \pm 0.042)$,
    marks the best-fit linear model. }
    \label{fig:B_P0} \, 
\end{figure}

%----------------------------------------
%\subsection{dependence of the cutoff period on the circularization parameter}
%\label{sec:param}
%-----------------------------------------
%
The relation between the cutoff period \Pcut\ and the circularization parameter $B$
is presented in Fig.~\ref{fig:B_P0} by a tight linear relation,  
$\log P_{\rm cut} = (0.198 \pm 0.011) \, \log (B) - (0.231 \pm 0.042)$,  
implying that  
\begin{equation}
    P_{\rm cut}  \varpropto B^{3/16} \, .
    \label{eq:B_P0}
\end{equation}
This can be traced back to equations~\ref{eq:tidal_approx} by noting that for small eccentricity the logarithmic derivative of $e$ is approximated by $\frac{1}{e}\frac{de}{dt}\propto B P^{-16/3}$. $P_{\rm cut}$ is the period for which $\frac{\Delta e}{e}\sim 1$ during the "lifetime" of the binary, so that 
$B P_{\rm cut}^{-16/3}\sim 1$.

%the logarithmic derivative of e $-\frac{1}{e} \frac{de}{dt}\sim 1$.

%--------------------------------------------------------------
%-----------------------------------------------------------
\newpage
%===================================
\section{Cutoff period vs.~effective temperature in MS binaries}
\label{sec:p0_temp}
\begin{figure}
	\includegraphics[width=8.5cm]{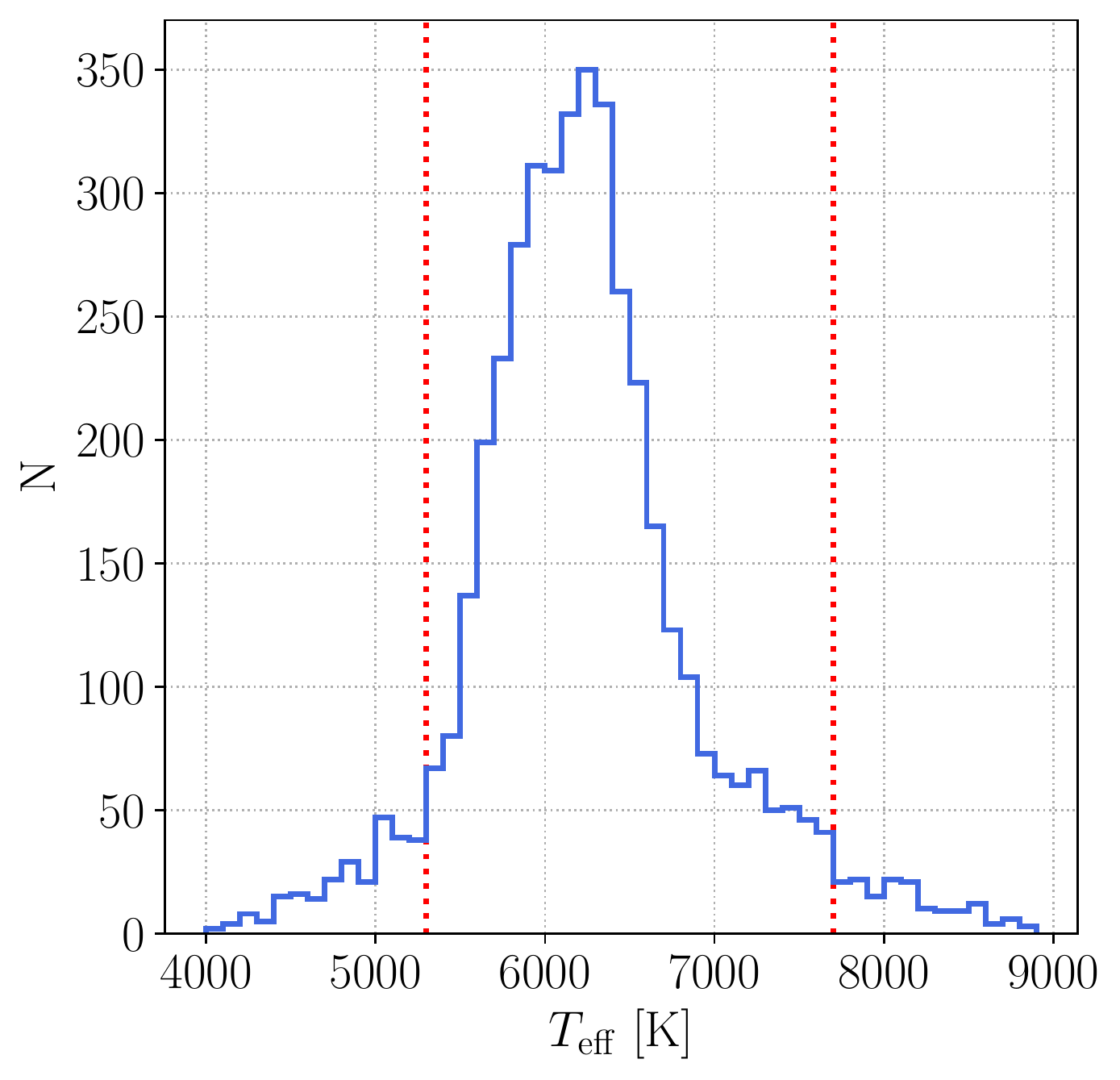}
    \caption{Stellar effective temperature histograms of our non-circular MS, $P < 30$ day, binary sample. Inside the range marked by red vertical dotted lines ($T_{\mathrm{eff}}=5300-7700$) there are $3\, 959$ binaries.}
    \label{fig:TeffHist}
\end{figure}

%===============================
The large number of \gaia MS binaries at hand enables us to look for the dependence of the cutoff period on the stellar temperature, which is readily available for most binaries. Fig.~\ref{fig:TeffHist} shows a histogram of stellar temperatures of the $4\,376$ orbits we have at hand (see above). We proceeded with the $3\, 959$ binaries in the range $5300$ -- $7700$ K.

%--------------------------------------------------------------
%--------------------------------------------------------------

\begin{figure*}
	\includegraphics[width=15cm]{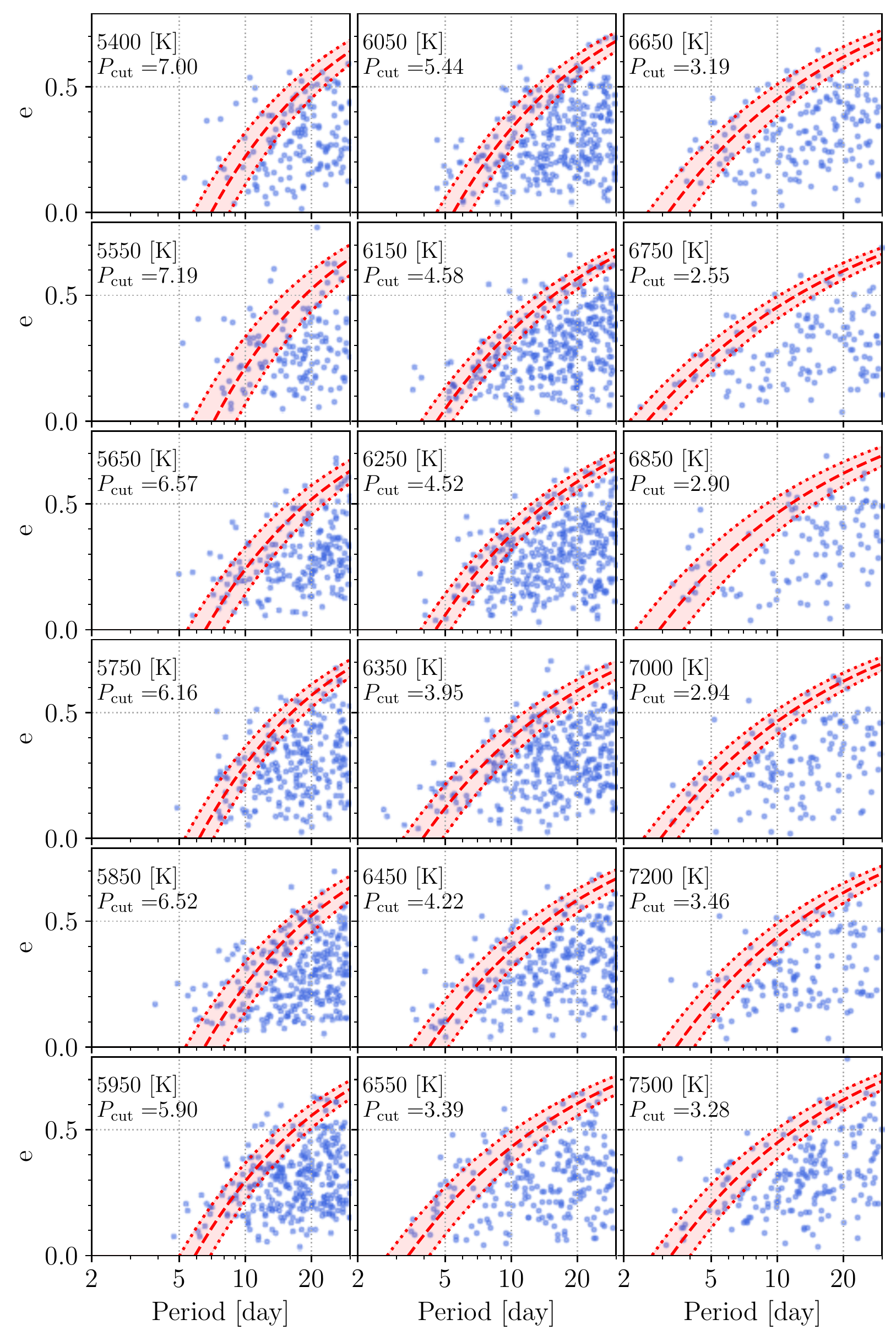}
    \caption{Period-eccentricity diagram with fitted upper envelope (see Table~\ref{tab:P0-Teff} for details) in red dashed line for the $18$ sub-samples of the MS \gaia binaries. The red area marks the transition region of $\pm \delta$ along the envelope. 
    }
    \label{fig:per_ecc_teff}
\end{figure*}

\begin{figure*}
	\includegraphics[width=13cm]{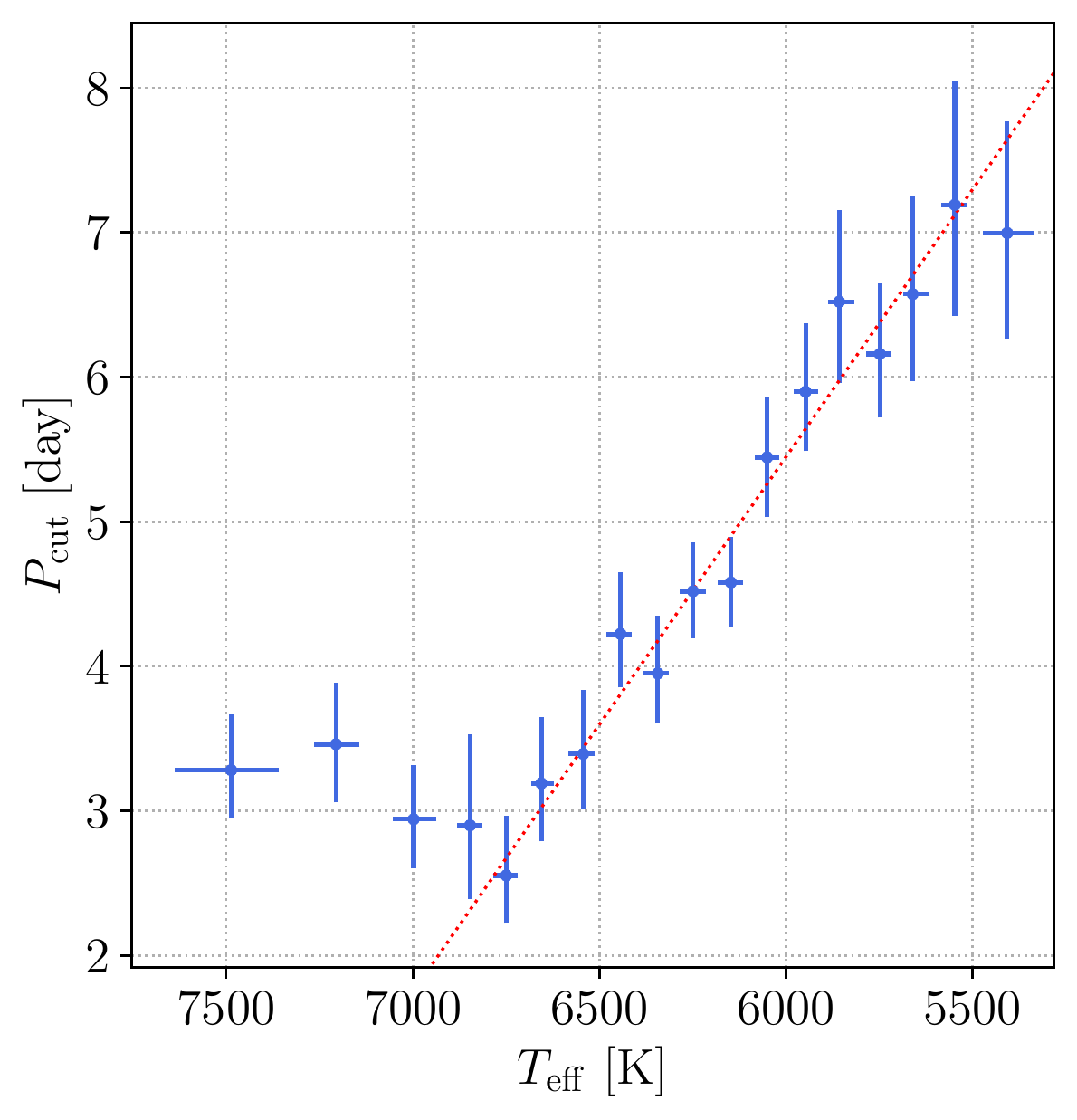}
    \caption{ \Pcut\ of $18$ \gaia sub-samples as a function of their corresponding temperature. Red-dotted line, 
    $P_{\rm cut}=-3.69\, (\pm 0.39)\,
          ({T_{\rm eff}/1000 {\rm K}}) +  27.6\, (\pm 2.5) {\rm \ day}$,   
    marks the best-fit linear model ignoring the rightmost and four leftmost bins.}
    \label{fig:teff_p0}
\end{figure*}

We divided those binaries into $18$ temperature bins and fitted an upper envelope to each sub-sample with the our algorithm.
%as in section~\ref{sec:envelope}. 
We show in Fig.~\ref{fig:per_ecc_teff} the upper envelope best-fit model (red curve) and transition region of $\pm \delta$ width along the envelope (red area) of these binary samples. The results are summarized in 
Table~\ref{tab:P0-Teff} 
and Fig.~\ref{fig:teff_p0}, where the points and error bars mark our median and $16\%,~84\%$ percentiles. 

%-----------------------------------------------------
\begin{table}
\caption{
Best-fitted values of \Pcut, $\tau$, $\delta$ for $18$ temperature bins of the \gaia MS spectroscopic binaries.}
%{\color{red} maybe the range of temperature}
%-------------------------------------
\begin{center}
\begin{tabular}{c c c c c }
\hline
\Teff~[k] & N & \Pcut~[day] & $\tau$ & $\delta$ \\
\hline
$5300-5500$ &$148$ & $7.00\pm 0.77$ & $1.42\pm 0.16$ & $0.08\pm 0.01$\\
$5500-5600$ &$137$ & $7.19\pm 0.86$ & $1.37\pm 0.17$ & $0.10\pm 0.02$\\
$5600-5700$ &$198$ & $6.57\pm 0.68$ & $1.52\pm 0.14$ & $0.08\pm 0.01$\\
$5700-5800$ &$237$ & $6.16\pm 0.49$ & $1.40\pm 0.11$ & $0.07\pm 0.01$\\
$5800-5900$ &$280$ & $6.52\pm 0.63$ & $1.52\pm 0.12$ & $0.09\pm 0.01$\\
$5900-6000$ &$310$ & $5.90\pm 0.47$ & $1.50\pm 0.11$ & $0.07\pm 0.01$\\
$6000-6100$ &$307$ & $5.44\pm 0.41$ & $1.49\pm 0.11$ & $0.08\pm 0.01$\\
$6100-6200$ &$334$ & $4.58\pm 0.32$ & $1.76\pm 0.11$ & $0.07\pm 0.01$\\
$6200-6300$ &$348$ & $4.52\pm 0.34$ & $1.67\pm 0.11$ & $0.07\pm 0.01$\\
$6300-6400$ &$336$ & $3.95\pm 0.40$ & $1.83\pm 0.13$ & $0.09\pm 0.01$\\
$6400-6500$ &$261$ & $4.22\pm 0.43$ & $1.77\pm 0.14$ & $0.09\pm 0.01$\\
$6500-6600$ &$222$ & $3.39\pm 0.44$ & $1.91\pm 0.17$ & $0.10\pm 0.02$\\
$6600-6700$ &$165$ & $3.19\pm 0.46$ & $1.91\pm 0.22$ & $0.09\pm 0.02$\\
$6700-6800$ &$123$ & $2.55\pm 0.41$ & $2.26\pm 0.23$ & $0.08\pm 0.03$\\
$6800-6900$ &$104$ & $2.90\pm 0.63$ & $1.98\pm 0.24$ & $0.11\pm 0.03$\\
$6900-7100$ &$136$ & $2.94\pm 0.38$ & $1.95\pm 0.20$ & $0.08\pm 0.02$\\
$7100-7300$ &$126$ & $3.46\pm 0.43$ & $1.84\pm 0.20$ & $0.08\pm 0.02$\\
$7300-7700$ &$187$ & $3.28\pm 0.38$ & $1.87\pm 0.16$ & $0.09\pm 0.01$\\
\hline
\end{tabular}
\end{center}
\label{tab:P0-Teff}
\end{table}
%----------------------------------

%--------------------------------------------------------------
%--------------------------------------------------------------

%--------------------------------------------------------------
%-----------------------------------------------------------

%\newpage

The figure and the table offer a few features.
\begin{itemize}
    \item[--] \Pcut\ drops linearity from 
    $\sim 6.5$ day to 
    $\sim 2.5$ day  when moving from \Teff\ of $5500$ to
    $6800$. 
    \item [--] The linear model for the $13$ bins is 
    \begin{equation}
        \frac{P_{\rm cut}}{\rm day}=
        A + B\left(\frac{T_{\rm eff}}{1000K}\right)\, ,
    \end{equation}
    %\textcolor{red}{
    where 
    $A=27.6\pm2.5$ and $B=-3.69\pm0.39$.
    \item [--] The linear fit is quite tight, suggesting that the uncertainties in \Pcut\ are overestimated by a factor of $\sim 1.7$. This could be expected, as the typical uncertainty in \Teff\ is of the same order as the bin width we used, inserting extra noise into the bins. 
    \item[--]\Pcut\ is flat at the 
    $7700$ -- $6800$ K range ($\sim 3$ day), and might be also flat at $5600$ -- $5300$ K.
    \item[--] A possible small jump of \Pcut\ might be seen at the \cite{kraft67} break of $6100$ K.
    \item[--] There is a strong correlation between \Pcut\ and $\tau$ of the 
    $18$ bins ---
    the shorter \Pcut\ the larger $\tau$ is. This reflects the fact that the shape of the different upper envelopes is such that all reach the same eccentricity at a period of $\sim30$ day. This can be explained by the fact that binaries with such relatively long periods retain their primordial eccentricities and therefore do not show any dependence on \Teff.
 
\end{itemize}

\newpage

%
%===============================
\subsection{Cutoff period dependence on stellar age?}
\label{sec:p0_temp_age}
%===============================

As shown in Fig. \ref{fig:AgeTeff}, the MS \gaia binaries show a strong correlation between stellar temperature and age. In this sub-section we examine the dependence of \Pcut\ on both stellar parameters and show that the dependence on age, is weak if existing at all.

\begin{figure}
	\includegraphics[width=8.5cm]{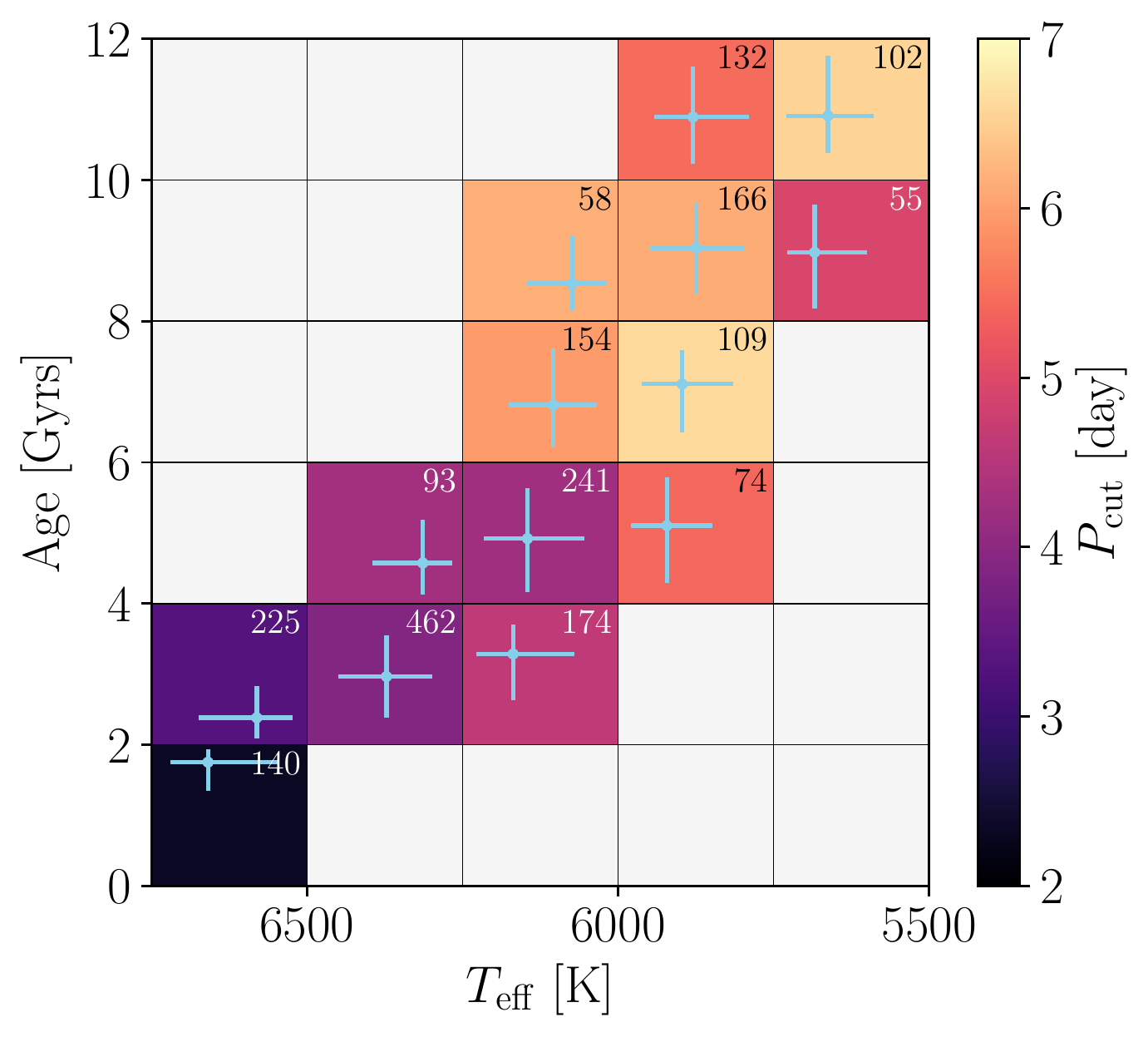}
    \caption{2D histogram  in evenly spaced bins of effective temperature and age. Colour represents \Pcut\ of \gaia binaries in bins with more than $50$ sources (otherwise white).
    Grey crosses mark median and $16\%,~84\%$ percentiles in each bin. The number of sources in each bin is displayed on the bin's top right side.}
    \label{fig:teff_age_P_cut bins} \, 
\end{figure}

We divided the age-temperature plane into evenly spaced bins of $250$ K and $2$ Gyr, and derived \Pcut\ for  the $14$ bins that contain more than $50$ binaries. These \Pcut\ values are colour coded in Fig.~\ref{fig:teff_age_P_cut bins}.

We fitted to the $14$ bins a two-dimensional function of
\begin{equation}
    \frac{P_{\rm cut}}{\rm day}=A_{\scriptscriptstyle\rm 2D} + B_{\scriptscriptstyle\rm 2D}\left(\frac{T_{\rm eff}}{1000K}\right) + C_{\scriptscriptstyle\rm 2D} \left(\frac{\mathrm{Age}}{ \mathrm{Gyr}}\right) \, ,
\end{equation}
and ran an MCMC procedure to find the best values.
We obtained 
$A_{\scriptscriptstyle\rm 2D}=25.5 \pm 6.9$, 
$B_{\scriptscriptstyle\rm 2D}=-3.46\pm 1.01$ and 
$C_{\scriptscriptstyle\rm 2D}=0.08 \pm 0.11$.

This shows that \Pcut\ strongly depends on stellar temperature and does not depend on the stellar age.

%--------------------------------------------------------------
%-----------------------------------------------------------

%===============================
\section{Approximated scaling law of circularization}
\label{sec:tidal-theory}
%===============================

%The previous section shows a clear dependence of the cutoff period 
%on the stellar temperature. 
%---  cooler stars display longer cutoff periods, which imply more effective circularization. 
%The sample can be divided into two parts --- hotter binaries with F-star and cooler ones with  G-star primaries. 

%
%According to common wisdom, \cite[e.g.,][]{kopal56,zahn75, zahn77, zahn89} a star in a binary system is subject to the gravitational tide of its companion, which induces a tidal bulge on its surface. 
%Unless the binary is circularized, synchronized, and aligned with the stellar rotation, the bulge does not point toward the companion and therefore exerts a tidal torque on the other star, which tends to bring the system to circularization, synchronization, and alignment.

In this section, we try to scale the tidal interaction with the stellar mass and radius and the orbital period, in order to derive the theoretical expectation for the dependence of \Pcut\ on the stellar temperature and compare it to our observational results.

According to \cite{zahn75,zahn77,zahnBouchet89,zahn89} and \cite{zahn08},
the timescale of the circularization is determined by the 
turbulent dissipation, in cool stars with convective envelopes, and radiative
damping, in hotter stars with radiative envelopes. 
%Thus, the equations that govern the two types of interaction are  different
We consider here the circularization of stars with radiative envelopes only, as at least half of the \gaia binaries analyzed here are of that type.

In a very simplified way, we consider only the eccentricity decay and ignore the period change associated with it. This can be close to reality only when the eccentricity is relatively small. Nevertheless, this approximation is probably good enough for the purpose of this section.
%
%\subsection{Dynamical tide of stars with radiative envelope}
%----------------------------------------
%\label{sec:RadiativeStars}
%
The approximation of the eccentricity derivative, to first order in $e$,  was already presented in equation~\ref{eq:tidal_approx}
%\citep[see][]{zahn75,zahn77,zahnBouchet89,zahn89, EylenWinnAlbrecht16} 
%can be written as
%---------------------------------------------------------------
\begin{equation}\tag{7}
\begin{aligned}
&-\frac{1}{e}\frac{de}{dt}= B_{\rm rad}{P^{-7}},\ \ 
%{\rm and} \\
%&\frac{dP}{dt}=-3e^2\,{B}{P^{-13/3}}\ ,
\end{aligned}
\label{eq:tidal-sim}
\end{equation}
%---------------------------------------------------------------
%
where we follow the convention of Section~\ref{sec:Simulations}.
%of equations~\ref{eq:tidal_approx}.
As emphasized there, changing the lifetime of a binary while keeping the circularization effectiveness is equivalent in our formulation to varying instead the $B_{\rm rad}$ parameter by the same factor.

Using \cite{ClaretCunha97} equation 18 \cite[see also][equation 3]{EylenWinnAlbrecht16}, we write
%
%-------------------------------------------
\begin{equation}
    B_{\rm rad}\propto 
    \frac{R_*^9}{M_*^3}\mathcal{F}(q)
    %\frac{q}{(1+q)^{5/3}}
    E_2\, T_{\rm MS}\, ,
\end{equation}
%-------------------------------------------------------
%
where $\mathcal{F}(q)=q/(1+q)^{5/3}$ is a function of the mass ratio, 
$E_2$ is a tidal constant evaluated numerically and $T_{\rm MS}$ is the MS lifetime. 
%and is a monotonic increasing function of the stellar mass \citep[see, for example, Fig.~1 of][]{ClaretCunha97}.
%
Adopting \cite{EylenWinnAlbrecht16} scaling of $R_*\propto M_*^{0.8}$ and $T_{\rm MS}\propto M_*^{-2.9}$, assuming the observed-stars age is a substantial fraction of their MS lifetime, we get 
%-------------------------------------------
\begin{equation}
    B_{\rm rad}\propto 
    {M_*^{1.3}}\mathcal{F}(q)
    E_2\,,
\end{equation}
%-------------------------------------------------------
%
Using equation 43 of \cite{hurley02},
$    E_2\propto     {M_*^{2.8}}$\,,
%-------------------------------------------------------
%
we finally get
%
%-------------------------------------------
\begin{equation}
    B_{\rm rad}\propto 
    {M_*^{4.1}}\mathcal{F}(q)
    \,,
\end{equation}
indicating that $B_{\rm rad}$ is a monotonic {\it increasing} function of stellar mass, and consequently of stellar temperature.

Following Fig.~\ref{fig:B_P0} and Equation \ref{eq:B_P0}, we conclude that for radiative-envelope primaries $P_{\rm cut} \varpropto B_{\rm rad}^{1/7}$
%We assume that the cutoff period can be approximated as the period for which 
%$B_{\rm rad}{P^{-7}}\sim 1$, ignoring the variation of $P$ itself, 
and therefore 
%
%the period for which The relation between the cutoff period \Pcut\ and the circularization parameter $B_{rad}$ is  %presented in Fig.~\ref{fig:simB_P0}. As expected
%---------------------------------------------------
\begin{equation}
    P_{\rm cut} \varpropto M_*^{0.6} \ .
\end{equation}
%---------------------------------------------------
%
This implies that \Pcut\ should also increase with stellar temperature, clearly in contrast with our results that show that \Pcut\ is {\it decreasing with stellar temperature (or mass)} in the range of $5500$ - $6700$ K.

%================================================
\section {Discussion}
\label{sec:discussion}
%================================================

We used the cleaned \gaia sample of spectroscopic binaries to study the distribution of the orbital eccentricity as a function of the binary period. The distribution is characterized by an upper envelope \citep{SB9,Mazeh2008}
that starts at \Pcut\ and eccentricity zero and rises monotonically toward high eccentricities for longer periods. The assumption is that circularization processes induced by tidal interaction between the two stars shaped the eccentricity distribution. The \gaia sample is a new opportunity to confront the theory of circularization with fresh data. 

We model the upper envelope by a simple function of two parameters, one of which is \Pcut, while  the other, $\tau$, determines the rise of the envelope.
We then follow the \cite{Mazeh16} approach and use
a modified Fermi function to describe a probability density distribution of the binaries above and below the upper envelope in the ($\log P$, $e$) plane.
The probability density function converges to zero above the envelope, and to a positive constant below it, with a transition region of a derived width.
We use an MCMC routine to find the best parameters of the distribution, given a sample of orbits.

The unprecedentedly large sample of \gaia orbits with MS A-, F- and G-type primaries enables us to derive the dependence of the envelope, and \Pcut\ in particular, on the stellar temperature in the range of
$5500$ -- $7500$ K.
To do that we divide the binaries into sub-samples of different temperatures, and fit the envelope of each sub-sample independently. 

%Obviously, each bin contains binaries with different ages and mass ratios and some temperature spread, as their uncertainties are probably on the order of 100 K. 

Our main finding is that \Pcut\ presents a  {\it linearly} decreasing function of stellar temperature for the G and F stars ---  from $6.5$ day for $5700$ K to $\sim 2.5$ day for $T_{\rm eff}\sim 6800$ K,  at a rate of $-3.7$ days/1000 K. The linear slope is at more than $9 \sigma$ significance. 

The uncertainties of the different \Pcut\ values are larger than the scatter around the linear fit. We suggest these are due to "imperfections" of the samples, the result of different primordial eccentricity and period distributions, binaries with different ages and mass ratios, and stellar temperature spread, as the $T_{\rm eff}$ uncertainties are probably on the order of 100 K. 
It is quite surprising that, nevertheless, the linear dependence of \Pcut\ is so pronounced and tight over a range of temperatures that correspond to stellar envelopes of convective and radiative nature alike.     

%Starting at the hottest stars at hand, \Pcut\ increases linearly  
In addition,  \Pcut\ is probably flat for \Teff\ at $5500$  --  $5700$ K, and $6800$  -- $7500$ K. At the well-known stellar Kraft brake of $\sim6100$ K \citep[see][]{kraft67} we possibly see a small jump, yet this is still barely significant.

We do not see any reason to believe our results are due to some \gaia observational bias. For example, we could not find any dependence of 
the quality score of \cite{bashi22}, 
which estimates the degree of validity of the orbits, on stellar temperatures or orbital periods and eccentricities.

As we have shown, the \Pcut\ trend revealed for the G- and F-stars 
%with radiative envelopes
%with $T_{\rm eff}\gtrsim 6100$, 
is inconsistent with Zahn's circularization theory, when we assume that circularization took place during the stellar MS lifetime.
This inconsistency could indicate that 
\begin{itemize}
    \item[--] Zahn dynamical tide theory is not accurate enough to account for the circularization processes. An assertion of studies suggested different approaches
    \citep[e.g.,][]{alexander73,hut81, tassoul87, hut92,dolginov92, GoldmanMazeh91, GoldmanMazeh94,goldman08, WitteSavonije99,WitteSavonije01,WitteSavonije02,
    DuhuidBaker20, ZanazziWu21, Terqem21,
    TerquemMartin21,
   koenigsberger21,zanazzi22,barker22,preece22,wei22}, 
    in stars with radiative envelopes in particular; see, for example, a heated discussion between \cite{tassoul97} and \cite{rieutord92} and \cite{ RieutordZahn97},  or
 %   {MathieuMazeh88}.   
%    \item The circularization is more effective for cooler stars with large stellar convective envelopes,     where the tidal dissipation takes place 
%    \citep[e.g.,][]{Albrecht12,EylenWinnAlbrecht16}.
    \item[--] The eccentricity distribution of the binaries was determined during the pre-main-sequence (PMS) phase of these stars, when the stars were much larger and therefore the circularization processes much faster, as suggested by \cite{MayorMermilliod84}, and worked out by \cite{zahnBouchet89} and later by 
    \citet[][KK11]{khaliullin11}, using updated PMS models; see also \cite{TerquemMartin21}.
\end{itemize}

The latter conjecture is interesting, especially because it is coming from Zahn himself. \cite{zahnBouchet89} claimed that for masses of $0.5$ -- $1.25 M_{\odot}$, the circularization took place during the PMS phase with expected \Pcut\ between $7.2$ and $8.5$ day \citep[see some observational evidence by][]{MathieuLathamMazeh92,mathieu94,melo01}.
%
% Although \cite{zahnBouchet89} study did not include the early-type stars considered here, one can assume that their arguments are relevant for a larger range of mass or temperature.
In fact, assuming the circularization occurred during the PMS phase, KK11 published a series of  expected \Pcut\ for different stellar masses based on updated PMS evolutionary tracks. Although the tracks are prone to a few uncertainties, the table of  KK11 
%\cite{khaliullin11}
displays a trend that might be similar to our results --- {\it \Pcut\ decreases as a function of the primary mass}.

%Resolving the apparent gap between the surfacing behaviour of \Pcut\ and the 
Understanding the theory of tidal circularization is crucial for understanding the evolution of short-period binaries \citep[e.g.,][]{hurley02,fragos22}. This has deep impact on how we think short-period binaries \citep[e.g.,][]{fabrycky07}, close triple systems \citep{MazehShaham79,naoz16AARA,toonen22}, cataclysmic binaries \citep[e.g.,][]{patterson84} and X-ray binaries \cite[e.g.,][]{bildsten92, podsi02} have evolved. Even some of the black-hole merger models depend on tidal interaction \cite[e.g.,][]{AntonininToonenHamers17,belczynski02}.

Finally, tidal interaction  could have a crucial role in the formation and evolution of exoplanets
\citep[][]{gu03,ida04,oglivie04,terquem07,jackson08, rao18}, hot Jupiters  \cite[e.g.,][]{rasio96,wu05,ferraz-mello08,leconte10,dawson18} and their orbit alignment with the stellar spin 
\citep{dobbes-dixon04,WinnHolman05,FabryckyWinn09,scherrer10,lai12,albrecht12,dawson14, mazeh15, Albrecht21,AlbrectDawsonWinn22}
in particular. 

In the future, five observational avenues might be of use in order to deepen our understanding of the tidal interaction in binary systems. First, one can use other samples of spectroscopic binaries, to confirm our results and to find the dependence of \Pcut\ on other parameters, like the mass ratio, stellar age and metallicity. Such samples include, for example, the next planned \gaia release,\footnote{https://www.cosmos.esa.int/web/gaia/release} and results of multi-object spectrographs, like RAVE \citep{RAVE11}, LAMOST \citep{cui2012large}, APOGEE \citep{Price-Whelan20} and the 
near-future 4MOST \citep{4MOST19}. 

The second avenue, analyzing samples of eclipsing binaries,  was already taken by \cite{NorthZahb03,MazehTamuzNorth06,EylenWinnAlbrecht16} and \cite{JustesenAlbrecht21}, 
as detailed in the introduction.
The advantage of using EBs is the capability to  accurately derive small eccentricities, which is not possible for SB1s. 
%
%On the other hand, the population of EBs is concentrated on short-period binaries, and EBs with periods larger than, say, 10 days, are quite less frequent. Thus, fitting a cutoff period or cutoff ratio by using the whole spectrum of periods is difficult.  Nevertheless, 
It would be interesting to use additional samples, like that of \gaia \citep{mowlavi22}
and derive \Pcut\ as a function of \Teff.
The large photometric surveys at work, like ZTF \citep{ZTF_EB_20} and ASAS-SN \citep{ASAS_EB_06,ASAS_EB_22} can yield large samples of additional eclipsing binaries.
%with new cutoff ratios. 

The third avenue has to do with the obvious realization that the long-term tidal interaction is not limited to circularization but acts to synchronize \citep[see][]{khaliullin10} and align the binary with the stellar rotation \citep[e.g.,][]{hurley02,naoz14}. 
Therefore, a sample of binaries should also display synchronization and alignment cutoff periods. It would be interesting if we could compare those periods  with the circularization period and their dependence on stellar parameters.

To follow the stellar rotation one can use the available large set of photometric light curves, like those of OGLE \citep[e.g.,][]{OGLE16}, {\it TESS} \citep[][]{TESS20} and \gaia \citep{eyer22}, which can reveal the stellar rotation periods \citep{mcquillan13,mcquillan14,avallone22}.  
Synchronization can be assumed, for example, if one detects ellipsoidal variability, which is modulated with the binary period  \citep{faigler11,faigler12,green22}. 
%Independent stellar rotation can be derived from the stellar photometric modulation  and be compared to the binary period for evidence for non-synchronized rotation. 

The fourth avenue has to do with observed eccentric
pseudosynchronized binaries \citep{hut81}, discovered in the {\it Kepler}
lightcurves \citep[e.g.,][]{zimmerman17,saio22}, which are going through strong tidal interaction. It would be interesting to compare their stellar rotation periods with the tidal theory expectation. In one of these systems, an orbital-period decay driven by tidal interaction has been probably observed \citep{WASP-12_Ou21}.

Finally, recently some observational evidence has been presented for the orbital decay of very close hot Jupiters, like WASP-12 \citep[e.g.,][]{WASP-12_YeeWinn20,WASP-12_turner21,WASP-12_WongShporerWinn22}; see also \cite{WASP-12_Harre22}, \cite{WASP-12_YangWei22} and \cite{WASP-12_rosario22}, based on the precise timings of transits that span over more than a decade. Following  the planetary period decay in real time
%before our eyes 
\citep{WASP-12_Yang22}
has the exciting potential to further constrain the theory of tidal interaction.

When additional information is available, 
the observational evidence could be compared with 
a more realistic tidal model that might be combined with some conjecture about the primordial period-eccentricity distribution. Such a model may include other evolutionary mechanisms, like magnetic breaking \citep[e.g.,][]{Mestel68, fleming19} and interaction with accretion discs and/or third companions \citep[e.g.,][]{fabrycky07,naoz16AARA,toonen22}, so it can account for the observed statistical features of the population of short-period binaries.

%The view presented here that considers the circularization by itself is meant to be only a first  simplistic step towards a deeper understanding of the tidal interaction between the stars and their companions.

%We, therefore, hope that the present study will initiate a combined theoretical and observational effort that will allow accounting for the results of the analysis of the \gaia binary sample. 

%\clearpage
%=================================================
\section*{Acknowledgements}
%=================================================

%
%\textbf{We wish to express our deep gratitude to the 
%referee for their thoughtful comments on the previous version of this paper. Following their wise advice, we were able to substantially improve the manuscript.}
%
We are deeply indebted to the {\it NSS} group and all the \gaia team for producing a vast high-quality catalogue that enabled us to follow the tidal circularizationof short-period binaries. We are extremely grateful to Josh N. Winn and Robert D. Mathieu for their illuminating comments and wise  suggestions that have significantly improved this manuscript.
The referee contributed very helpful comments on a previous version of the paper, helping us presenting our results in a clearer way.
This research was supported by Grant No. 2016069 of the United States-Israel Binational Science Foundation (BSF) to TM, 
and
Grant No. I-1498-303.7/2019 of the German-Israeli Foundation for Scientific Research and Development (GIF) to TM.  
This work has also made use of data from the European Space Agency (ESA) mission \textit{Gaia} (https://www.cosmos.esa.int/gaia), processed by the \textit{Gaia} Data Processing and Analysis Consortium (DPAC; https://www.cosmos.esa.int/web/gaia/
dpac/consortium). Funding for DPAC has been provided by national institutions, in particular the institutions participating in the \textit{Gaia} Multilateral Agreement.

%This research made use of \texttt{exoplanet} \citep{foreman21} and its dependencies \citep{astropy13,astropy18, salvatier2016, team2016theano, luger2019starry, agol2020analytic}. %exoplanet:pymc3, exoplanet:theano}

%%%%%%%%%%%%%%%%%%%%%%%%%%%%%%%%%%%%%%%%%%%%%%%%%%
\section*{Data Availability}

Data used in this study are available upon request from the corresponding
author.



\bibliographystyle{mnras}
\bibliography{MNRASbib} % if your bibtex file is called example.bib

% Don't change these lines
\bsp	% typesetting comment
\label{lastpage}

\end{document}